\documentclass[prb,showpacs,showkeys,aps,amsmath,groupedaddress,eqsecnum]{revtex4}

\usepackage{amssymb}
\usepackage{amsmath}
\usepackage{graphicx}
\usepackage{amsbsy}
\usepackage{color}

\newcommand{\bq}{\begin{eqnarray}}
\newcommand{\eq}{\end{eqnarray}}
\newcommand{\bqn}{\begin{eqnarray*}}
\newcommand{\eqn}{\end{eqnarray*}}
\newcommand{\nn}{{\bf n}}
\newcommand{\rr}{{\bf r}}

\begin{document}
\title{Cluster theory of Janus particles} 

\author{Riccardo Fantoni}
\email{rfantoni27@sun.ac.za}
\affiliation{National Institute for Theoretical Physics (NITheP) and
Institute of Theoretical Physics, Stellenbosch University,
Stellenbosch 7600, South Africa} 
\author{Achille Giacometti}
\email{achille@unive.it}
\affiliation{Dipartimento di Chimica Fisica, Universit\`a di Venezia,
S. Marta DD2137, I-30123 Venezia, Italy}
\author{Francesco Sciortino}
\email{francesco.sciortino@phys.uniroma1.it}
\affiliation{Dipartimento di Fisica, Universit\`a di Roma La Sapienza,
Piazzale A. Moro 2, 00185 Roma, Italy}
\author{Giorgio Pastore}
\email{pastore@ts.infn.it}
\affiliation{Dipartimento di Fisica dell'
Universit\`a, Strada Costiera 11, 34151 Trieste, Italy}  
\date{\today}

\begin{abstract}
We apply a simple statistical mechanics cluster approximation for
studying clustering in the Kern and  Frenkel model of Janus
fluids. The approach is motivated by recent Monte Carlo simulations work on the same model 
revealing that the vapor coexisting with the liquid phase contains
clusters of different sizes and shapes whose equilibrium concentrations in general
depend on the interaction range as well as on thermodynamic
parameters. The approximation hinges on a separation between the intra- and inter-cluster contribution to thermodynamics, 
where only the former is explicitly computed by Monte Carlo simulations. Two levels of a simple liquid theory approximation are exploited 
for the description of the latter. In the first we
use the ideal-gas expressions and obtain a qualitative agreement with extensive Monte Carlo bulk simulations.
This can be improved to a semi-quantitative agreement, by using a hard-sphere description for the cluster-cluster correlations.
\end{abstract}

\maketitle
\section{Introduction}

\label{sec:0}
Recent advances in experimental techniques for chemical sysntesis have provided
a well defined set of different protocols for obtaining colloidal particles with
different shapes, chemical compositions and surface patterns. In particular it is now possible 
to obtain colloidal particles with pre-defined number and distribution of solvophobic and solvophilic
regions on their surface. These are usually referred to as patchy colloids.\cite{Manoharan2003,Pawar2010,Glotzer2007,Zhang2004}

The simplest example within this realm is constituted by the so-called Janus particles, where the
surface is partioned in only two parts with even distribution of the two philicities.
In spite of their apparent simplicity, Janus particles have aroused increasingly interest in the last few years both
for their potential technological applications and in view of the rather unusual displayed self-assembly properties as compared to
conventional isotropic colloidal particles.  \cite{deGennes1992,Casagrande1989,Hong2006,Walther2008}.

A detailed study of the fluid-fluid transition for Janus fluids has recently been carried out by Monte Carlo (MC) simulation 
\cite{Sciortino2009,Sciortino2010} using the Kern-Frenkel pair potential \cite{Kern03}. 
Within this model, the solvophobic and solvophilic hemispheres are mimicked
by an attractive square-well potential and a repulsive hard-sphere potential respectively, and two spheres attract each other only
provided that their centers are within a given distance, as dictated by the width of the well, and the two attractive patches on each sphere 
are properly aligned one another, that is lie within a predefined relative angular range.

The combined features of the equal amplitude of the two philicities coupled with the specificity of the chosen potential types, give rise to
a micellization process originating in the vapor phase that severely destabilizes the condensation process thus providing a re-entrant 
vapour coexistence curve that in the temperature-density diagram is skewed toward higher densities as the system is cooled to lower 
temperatures \cite{Sciortino2009}. A number of additional unusual features were also found for the vapour phase \cite{Sciortino2010},
including the fact that, for the chosen width of the square-well potential ($50\%$ of the particle size), there appeared a predominance of
particular clusters formed by single-layer (micelles of about 10 particle)  and double-layer (vesicles of about 40 particles) always
exposing the hard-sphere part as an external global surface, thus inhibiting the formation of a liquid phase.

It should be emphasized that MC simulations are particularly demanding for this system in such that very low temperatures (of the order of
0.25 or less in reduced units) are necessary to observe these phenomena, and this is expected to be even more demanding for decreasing
range of the interactions \cite{Sciortino2010}.

In this paper, we focus our interest to the study of the vapor phase, following  a different approach, hinging on a strategy similar to that devised in the context of associating fluids, where
several different theories with different degrees of success have been envisioned. \cite{Wertheim-ass,Evans1997,Fisher1998,Freed2008} 

Our approach has been inspired by the work of Tani and Henderson \cite{Tani83}, extending the Bjerrum theory for association
in electrolytic solutions \cite{Bjerrum26} where the total partition function
is factorized into a intra- and inter-cluster contribution, so that
the original task is reduced to the computation of the partition
function for clusters of increasing sizes along with the interaction among them. 

While the original approach \cite{Tani83} was limited by the necessity of evaluating analytically even the intra-cluster 
partition function, in addition to the inter-cluster contribution, we propose to determine the former by explicit Monte Carlo simulations 
for each cluster and the latter using physically motivated
fluid theories. Within MC simulation of each $n-$cluster, we are then able to determine the energy per particle
as a function of temperature and thereby compute the
excess free energy of the isolated cluster by thermodynamic integration.

Clearly, this approach is particularly suited to study  the vapor  
phase as once the first n-particle clusters have been simulated the  
resulting information can be inserted in the inter-cluster theory,
and this is enough to determine the partition function of the vapor at  
all thermodynamic states.
We can then follow and predict the dependence of cluster population on  
thermodynamic conditions and interaction parameters. This is  
particularly relevant in cases in which spontaneous
cluster formation is particularly slow, due for example to the low  
value of the temperature at which clustering takes place, a common case 
when the  
interaction range is very short.
 
The paper is organized as follows: in section \ref{sec:1} we describe
the model, in sections \ref{sec:2} and \ref{sec:3} we introduce the
cluster theory, in 
section \ref{sec:4} we describe how we determined the
intra-cluster partition function. Additional results are then presented in section \ref{sec:5}, and section \ref{sec:6} is for final remarks. 

\section{The Kern and Frenkel model}
\label{sec:1}
As in the work of Sciortino {\sl et. al.}
\cite{Sciortino2009,Sciortino2010} we used 
the Kern and Frenkel \cite{Kern03} patchy hard sphere model to
describe the Janus fluid. Two spherical particles attract via a
short-range square-well 
potential only if the line segment joining the centers of the two spheres
intercepts a patch on the surface of the first particle and one on the
surface of the other. In the case of a {\sl single} patch per
particle, the pair potential reads \cite{Kern03} 
\begin{eqnarray}
\label{kf:eq1}
\Phi(1,2)&=&\phi(r_{12})\Psi(\hat{\nn}_1,\hat{\nn}_2,\hat{\rr}_{12})~,
\end{eqnarray}
where
\begin{equation}
\label{kf:eq2}
\phi(r)=\left\{
\begin{array}{ll}
+\infty   & r<\sigma\\
-\epsilon & \sigma<r<\lambda\sigma\\
0         & \lambda\sigma<r
\end{array}\right.
\end{equation}
and
\begin{equation}
\label{kf:eq3}
\Psi(\hat{\nn}_1,\hat{\nn}_2,\hat{\rr}_{12})=\left\{
\begin{array}{ll}
1 & \mbox{if $\hat{\nn}_1\cdot\hat{\rr}_{12}\ge\cos\theta_0$ and 
$-\hat{\nn}_2\cdot\hat{\rr}_{12}\ge\cos\theta
_0$}\\
0 & \mbox{otherwise}
\end{array}\right.
\end{equation}
where $\theta_0$ is the angular semi-amplitude of the patch. Here
$\hat{\nn}_1(\omega_1)$ and $\hat{\nn}_2(\omega_2)$ are unit vectors
giving the directions of the center of the patch in spheres 1 and 2,
respectively, with $\omega_1=(\theta_1,\varphi_1)$ and
$\omega_2=(\theta_2,\varphi_2)$ their corresponding spherical angles
in an arbitrary oriented coordinate frame. Similarly,
$\hat{\rr}_{12}(\Omega)$ is the unit vector of the separation $r_{12}$
between the centers of the two spheres and is defined by the
spherical angle $\Omega$. As usual, we have denoted with $\sigma$ 
the hard core diameter and $\lambda=1+\Delta/\sigma$ with $\Delta$ the
width of the well.

One can define the fraction of surface covered by the attractive patch
as
\begin{eqnarray}
\label{kf:eq4}
\chi=\langle\Psi(\hat{\nn}_1,\hat{\nn}_2,\hat{\rr}_{12})
\rangle_{\omega_1,\omega_2}^{1/2}
=\sin^2\left(\frac{\theta_0}{2}\right)~.
\end{eqnarray} 

where we have introduced $\langle \ldots \rangle_{\omega} = (1/4 \pi) \int d \omega (\ldots) $ as the average over the solid angle $\omega$.  

Reduced units $k_B T/\epsilon$ ($k_B$ is the Boltzmann constant) and $\rho \sigma^3$ will be used as a measure of the temperature 
and density in numerical data. 


\section{A cluster theory for Janus particles}
\label{sec:2}
Following Ref. \onlinecite{Tani83}, we split the partition function in
an inter- and intra-cluster 
contribution. Let $N_n$ be the number of clusters formed by $n$
particles, where $n=1,\ldots,n_c$ ($n_c$ being the number of different
clusters) and $\rho_n=N_n/V$ their density. We then write the total
partition function as 
\begin{eqnarray}
\label{cluster:eq1}
Q_{\text{tot}}&=&\sum_{\{N_n\}}^{\prime} \left[\prod_{n=1}^{n_c}
\frac{1}{N_n!} \left(q_n^{\text{intra}}\right)^{N_n} \right] Q_{\text{inter}}
\left(\{N_n\},V,T\right)~, 
\end{eqnarray}
where the prime indicates that the sum is restricted to all possible
configurations satisfying the obvious constraint of conserving the 
total number of particles $N$,
\begin{eqnarray}
\label{cluster:eq2}
\sum_{n=1}^{n_c} n N_n &=& N~.
\end{eqnarray}
Here $q_n^{\text{intra}}$ is the ``internal'' partition function for a
$n-$particle cluster and $ Q_{\text{inter}} \left(\{N_n\},V,T\right)$ is
the inter-cluster partition function. Additional controlled thermodynamic variables
are the total volume $V$ and the temperature $T$.

The constraint can be dealt with by introducing a Lagrange multiplier
so that we minimize the quantity 
\begin{eqnarray}
\label{cluster:eq3}
\ln \widehat{Q}_{\text{tot}} &=& \ln Q_{\text{tot}} + \left(\ln \lambda\right)
\sum_{n=1}^{n_c} n N_n~.   
\end{eqnarray}
In computing the partition function (\ref{cluster:eq1}) we assume that
the sum can be replaced by its largest dominant contribution. With the 
help of Stirling approximation $N!\approx (N/e)^N$ one then obtains
\begin{eqnarray}
\label{cluster:eq4}
\ln Q_{\text{tot}}&\approx& \sum_{n=1}^{n_c} \left[N_n \ln
q_n^{\text{intra}} - \left(N_n \ln N_n - N_n\right)\right] + \ln
Q_{\text{inter}}~.
\end{eqnarray}
The correct cluster distribution $\{\overline{N}_n\}$ is then found
from the variational condition 
\begin{eqnarray}
\label{cluster:eq5}
\left.\frac{\partial}{\partial N_n} \ln \widehat{Q}_{\text{tot}}
\right|_{\{N_n=\overline{N}_n\}} &=& 0~. 
\end{eqnarray}
This allows the calculation of the resulting free energy, $\beta
F_{\text{tot}}=- \ln Q_{\text{tot}}$, in terms of the internal reduced
free energy densities, 
$\beta f_{n}^{\text{intra}}=-\ln q_n^{\text{intra}}$, so that
\begin{eqnarray}
\label{cluster:eq6}
\frac{\beta F_{\text{tot}}}{V} &=& \sum_{n=1}^{n_c}
\left[\overline{\rho}_n \ln \overline{\rho}_n - \overline{\rho}_n
\right] + 
\sum_{n=1}^{n_c} \overline{\rho}_n \beta f_{n}^{\text{intra}} +
\sum_{n=1}^{n_c} \overline{\rho}_n \ln V - \frac{1}{V} \ln
Q_{\text{inter}}~.
\end{eqnarray}
In the above expression, $\beta=1/(k_B T)$. 

\section{Specific models}
\label{sec:3}
We now consider two specific cases, where the inter-cluster  
interaction is not accounted for (the ideal-gas case) or modeled as  
an effective hard-sphere like interaction between clusters.
\subsection{Ideal gas}
\label{subsec:ideal}
The simplest possibility corresponds to consider different clusters as
non-interacting ideal particles so that 
\begin{eqnarray}
\label{ideal:eq1}
Q_{\text{inter}}\left(\{N_n\},V,T\right) &=&\prod_{n=1}^{n_c}
\left(\frac{V}{\Lambda_n^3}\right)^{N_n} \equiv Q_{\text{ideal}}~, 
\end{eqnarray}
where $\Lambda_n$ is the de Broglie thermal wavelength associated with
each $n-$cluster.    
 
Using Eqns. (\ref{cluster:eq5}) and (\ref{ideal:eq1}) one easily obtains
\begin{eqnarray}
\label{ideal:eq2}
\overline{\rho}_n &=& \lambda^n \frac{q_n^{\text{intra}}}{\Lambda_n^3}~,
\end{eqnarray}
where $\overline{\rho}_n=\overline{N}_n/V$.

The actual value of the Lagrange multiplier $\lambda$ can then be
numerically obtained upon inverting the constraint (\ref{cluster:eq2}) 
\begin{eqnarray}
\sum_{n=1}^{n_c} n \lambda^n \frac{q_n^{\text{intra}}}{\Lambda_n^3} &=&
\rho \equiv \frac{N}{V}~. 
\label{ideal:eq3}
\end{eqnarray}

Substitution of Eq. (\ref{ideal:eq1}) into the general expression of
the free energy (\ref{cluster:eq6}) leads to \cite{Barrat-note}
\begin{eqnarray}
\label{ideal:eq4}
\frac{\beta F_{\text{tot}}}{V} &=& \sum_{n=1}^N
\left[\overline{\rho}_n \ln \left(\overline{\rho}_n
\Lambda_n^3\right) - \overline{\rho}_n \right] 
+ \sum_{n=1}^N \overline{\rho}_n \beta f_{n}^{\text{intra}}~.
\end{eqnarray} 
\subsection{Chemical equilibrium}
\label{subsec:chemical}
The above result (\ref{ideal:eq4}) can be used to compute chemical equilibrium among different
clusters. Indeed, on defining $\mu_n$ as the chemical potential associated to the $n-$th cluster, we have
\begin{eqnarray}
\label{chemical:eq1}
\beta \mu_n = \frac{\partial\left(\beta F_{\text{tot}}\right) }{\partial N_n} = 
\frac{\partial\left(\beta F_{\text{tot}}/V\right) }{\partial \rho_n} &=& \ln \left(\rho_n \Lambda_n^3\right) + \beta f_{n}^{\text{intra}}
\end{eqnarray}
We can then impose the equilibrium condition $\mu_n=n \mu_1$ to obtain
\begin{eqnarray}
\label{chemical:eq2}
f_{n}^{\text{intra}} &=& n f_{1}^{\text{intra}} + k_B T \ln \left[ \frac{\rho_n \Lambda_n^3}{\left(\rho_1 \Lambda_1^3\right)^n} \right]
\end{eqnarray}
which can be used to compute the internal free energies, given the cluster distributions. An alternative procedure, based on the explicit
computation of the internal energy per particle within each cluster, will be discussed in Section \ref{sec:3}.

\subsection{Connection with Wertheim association theory}
\label{subsec:wertheim}
An interesting comparison can be found with Wertheim first-order association theory \cite{Wertheim-ass} which
is frequently used in this context (see e.g. Ref.\onlinecite{Sciortino2007} and references therein).
Within this theory, the bond contribution to the Hemholtz free energy  can be computed from a chemical equlibrium equation
under the condition that only a suitable subset of diagrams are included in the cluster expansion and each attractive site is
engaged at most in a single bond, the limit of a single-bond per patch in the language of the present paper.

Consider a system formed by only monomers and dimers, that is $n=1,2$. Then from Eq.(\ref{ideal:eq2}) and condition
(\ref{cluster:eq2}) limited to $n=1,2$ we can obtain a quadratic equation in the Lagrange multiplier $\lambda$. 
The only acceptable root can then be substituted into Eq.(\ref{ideal:eq2}) for $n=1$ to obtain
the fraction of patches that are not bonded, that is the fraction of monomers 
\bq
\label{wertheim:eq1}
\frac{\overline{\rho}_1}{\rho}=\frac{2}{1+\sqrt{1+8\rho \overline{\Delta}}}~,
\eq 
Using numerical simulations for $n=1,2$ clusters we are able to determine the energy per particle
in an $n-$cluster as a function of temperature and thereby determine the
excess free energy of the isolated cluster by integration
where 
$\overline{\Delta}=(q_2^{\text{intra}}/[q_1^{\text{intra}}]^2)(\Lambda_1^2/\Lambda_2)^3$. This equation is identical to the result from
Wertheim's theory (see Eq.(10) in Ref.\onlinecite{Sciortino2007}) when translated in the appropriate language. 
Therefore, the present formulation is equivalent to Wertheim's theory
provided that temperatures are sufficiently low (see Ref. \onlinecite{Sciortino2007} for further details) and 
the condition single-bond per binding site is satisfied.
On the other hand, the present theory allows for an arbitrary amplitude of
the patch thus including the possibility of multiple bonding.

Note that while in the case of only two clusters ($n=1,2$) requires the solution of a system of $2$ coupled equations that results
into a quadratic equation for $\lambda$, a general case with clusters up to the total number of clusters $n_c$ clearly requires the solution
of a system of $n_c$ coupled equation, a task that -- in general -- has to be carried out numerically.

\subsection{Effective hard sphere inter-cluster interaction}
\label{subsec:cs}
While simple, the ideal gas is clearly rather unphysical even at very
low densities. A more physical description amounts to consider all
$n-$particle clusters as identical hard spheres with diameters
$\sigma_n$ and packing fractions $\eta_n= (\pi/6) \rho_n \sigma_n^3$. 
A rather precise approximate solution in this case is provided by the
Boubl\'ik, Mansoori, Carnahan, and Starling expression
\cite{Boublik70,Mansoori71}, but for simplicity we here only consider
the case $\sigma_n=\sigma_0$ for all $n$, whose thermodynamics is well described by the simple
monodisperse Carnahan-Starling formulae \cite{Carnahan69}. 
This can be motivated by the fact that only a minor variation is found in the linear cluster dimensions (see Table \ref{tab:1}
and discussion further below) and by the observation that instantaneous size variations of an $n-$particle cluster are comparable with the variation of the average cluser radii for n within a few tens. It is then attempting to approximate the correlations between different shaped populations of clusters by a single effective one-component hard sphere system to take care of the average inter-cluster correlations.

Then
\begin{eqnarray}
\label{cs:eq1}
Q_{\text{inter}}\left(\{N_n\},V,T\right) &=& Q_{\text{ideal}} e^{- \beta
F_{\text{cs}}}~, 
\end{eqnarray}
where $Q_{\text{ideal}}$ is given in Eq. (\ref{ideal:eq1}) and
$F_{\text{cs}}$ is the Carnahan-Starling \cite{HanMcD86} excess free
energy 

\begin{eqnarray}
\label{cs:eq2}
\frac{\beta F_{\text{cs}}\left(\eta_t\right)}{N_{t}} &=& \frac{\eta_{t}
\left(4 -3 \eta_{t}\right)}{\left(1-\eta_t\right)^2}~, 
\end{eqnarray}

where $N_{t}=\sum_{n=1}^{n_c} N_n$ is the total number of clusters and
$\eta_{t}=\sum_{n=1}^{n_c}\eta_n$ is the total cluster packing fraction.  
    
Following the same steps as before one obtains
\begin{eqnarray}
\label{cs:eq3}
\overline{\rho}_n&=& \lambda^n 
\frac{q_n^{\text{intra}}}{\Lambda_n^3}G\left(\eta_{t}\right)~, 
\end{eqnarray}
where we have introduced the function
\begin{eqnarray}
\label{cs:eq4}
G\left(x\right) &=& \exp \left[-
\frac{x\left(8-9x+3x^2\right)}{\left(1-x\right)^3}\right]~. 
\end{eqnarray}
For the free energy one obtains from Eq. (\ref{cluster:eq6})
\begin{eqnarray}
\label{cs:eq5}
\frac{\beta F_{\text{tot}}}{V} &=& \sum_{n=1}^{n_c}
\left[\overline{\rho}_n \ln \left(\overline{\rho}_n
\Lambda_n^3\right) - \overline{\rho}_n \right] 
+ \sum_{n=1}^{n_c} \overline{\rho}_n \beta f_{n}^{\text{intra}} +
\frac{\beta F_{\text{cs}}\left(\overline{\eta}_{t}\right)}{V}~,
\end{eqnarray}
that differs from the ideal gas counterpart Eq. (\ref{ideal:eq4}) only
for the last additional term. Clearly one recovers the ideal gas in
the limit $\overline{\rho}_n \to 0$ as it should. In order to find the
correct solution for this system of equation it is important to choose
the one that is continuously obtained from the solution of the ideal
gas case at $\sigma_0\to 0$.

\subsection{Thermodynamic quantities}
\label{sec:thermodynamic}
It proves convenient to express our analysis in terms of reduced 
partition functions $Z$ rather than of the full partition functions
$Q$ used in Section \ref{sec:2}. This can be conveniently done by the
definitions 
\begin{eqnarray}
\label{development:eq1}
Q_{\text{inter}} =
\prod_{n=1}^{n_c}\frac{Z_{\text{inter}}}{\Lambda_n^{3N_n}}\qquad
q_n^{\text{intra}}=\Lambda_n^3Z_n^{\text{intra}}~.
\end{eqnarray}

Given the partition function $Q_{\text{tot}}$ 
we can determine the
Carnahan-Starling excess free energy 

\begin{eqnarray}
\label{therm:eq1}
\beta F^{exc}&=&-\ln\left(\frac{Q_{\text{tot}}}{V^{N}}\right)~,
\end{eqnarray}

the internal energy per particle

\begin{eqnarray}
\label{therm:eq2}
u=\frac{3}{2\beta}+\frac{1}{N}\frac{\partial(\beta F^{exc})}{\partial\beta}
=\frac{3}{2\beta}-\sum_{n=1}^{n_c}\frac{N_n}{N}
\frac{\partial\left(\ln Z_n^{\text{intra}}\right)}{\partial\beta}
=\frac{3}{2\beta}+\sum_{n=1}^{n_c} n\frac{N_n}{N}u_n(T)~,
\end{eqnarray}
where $u_n$ is the internal energy per particle of an $n-$cluster (see
Section \ref{sec:3}). We can also determine the compressibility factor
\begin{eqnarray}
\label{therm:eq3}
\frac{\beta P}{\rho}=\frac{1}{\rho}\frac{\partial\left(\ln
Q_{\text{tot}}\right)}{\partial V}= 
\frac{1}{\rho}\frac{\partial\left(\ln Z_{\text{inter}}\right)}{\partial V}=
\frac{1+\eta_{t}+\eta_{t}^2-\eta_{t}^3}{(1-\eta_{t})^3}~.
\end{eqnarray}

\section{Computation of the intra-cluster free energy}
\label{sec:4}
The simulation were carried out following the same prescription used  for the bulk fluid phases \cite{Sciortino2009,Sciortino2010}.
Two kind of moves for each chosen particle -- a random translation and a random rotation -- were allowed,  following standard recipes \cite{Allen}
and a standard Metropolis \cite{Metropolis53} algorithm was used to compute the energy per particle of the system of $n$ particles.

Typical runs were of about $5 \times 10^{6}$ steps, one step consisting of $n$ particles moves. 
 
We studied first the case of clusters in the neighborhood of $n=10$
particles which is expected to be sufficient to observe the micellization process due to the single
layer clustering \cite{Sciortino2010}.

To this aim we started with an initial configuration of two
pentagons with particles at their vertices juxtaposed one above the
other. The two pentagons are parallel to the $x-y$ plane, have the
$z$ axis passing from their centers, and are one at $z=+\sigma/2$ and
the other at $z=-\sigma/2$. The unit vectors attached to the spheres
where chosen so to connect the origin to the center of the given
sphere. We obtained the clusters with a lower
number of particles by simply deleting particles and obtained the
clusters with a higher number of particles by adding on the $z$ axis a
particle just above the upper pentagon and/or just below the lower
one. However the results of the simulations are independent of the
initial configuration chosen.

In order to compare with previous studies \cite{Sciortino2009}, we consider the $\Delta=0.5\sigma$ case first.

We performed the simulations of the isolated cluster and we have explicitly tested that results coincide
with the calculation stemming for the bulk low density Janus fluid from which we extract cluster informations by
taking all the cluster found with the same size and averaging their
properties. 

During the simulation we allow all possible moves but we do not count
the configurations which are not 
topologically connected, i.e. those configurations where it is not
possible to go from one sphere to all the others through a path; the
path being allowed or not to move from one particle 1 to particle 2
depending whether $\Phi(12)$ has value $-\epsilon$ or not.


At high temperatures the limiting value for the energy per particle is
$-\epsilon (n-1)/n$. At low temperature ($ k_B T/\sigma <0.15$) the clusters tend to
freeze into certain energy minima. This can be improved by ``regularizing'' the angular part 
of the Kern-Frenkel potential into
\begin{eqnarray}
\label{computation:eq1}
\Psi(\hat{\nn}_1,\hat{\nn}_2,\hat{\rr}_{12})=
\{\tanh[l(\hat{\nn}_1\cdot\hat{\rr}_{12}-\cos\theta_0)]+1\}
\{\tanh[l(-\hat{\nn}_2\cdot\hat{\rr}_{12}-\cos\theta_0)]+1\}/4~.
\end{eqnarray}
and  gradually increase $l$ starting from $1/2$ during the
simulation up to values where there is no actual difference between
the continuous potential and the original stepwise one.
This allowed us to reach the configuration with the real minimum
energy with a certain confidence. 

\begin{figure}[htbp] 
\centering
\includegraphics[width=3.5in]{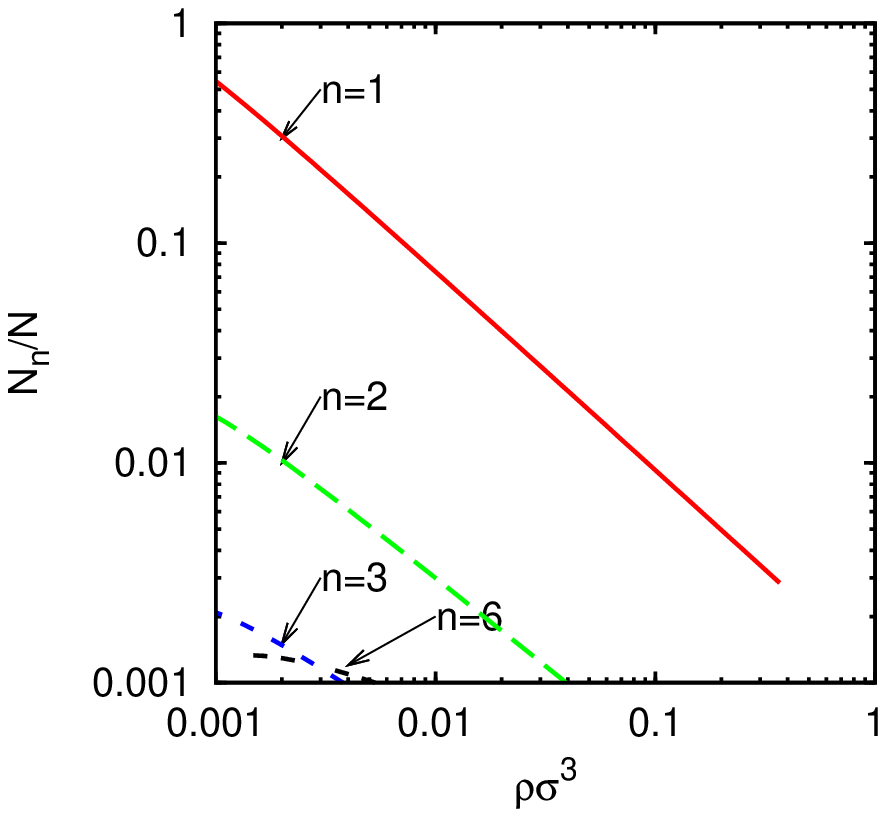}
\hfill 
\includegraphics[width=3.5in]{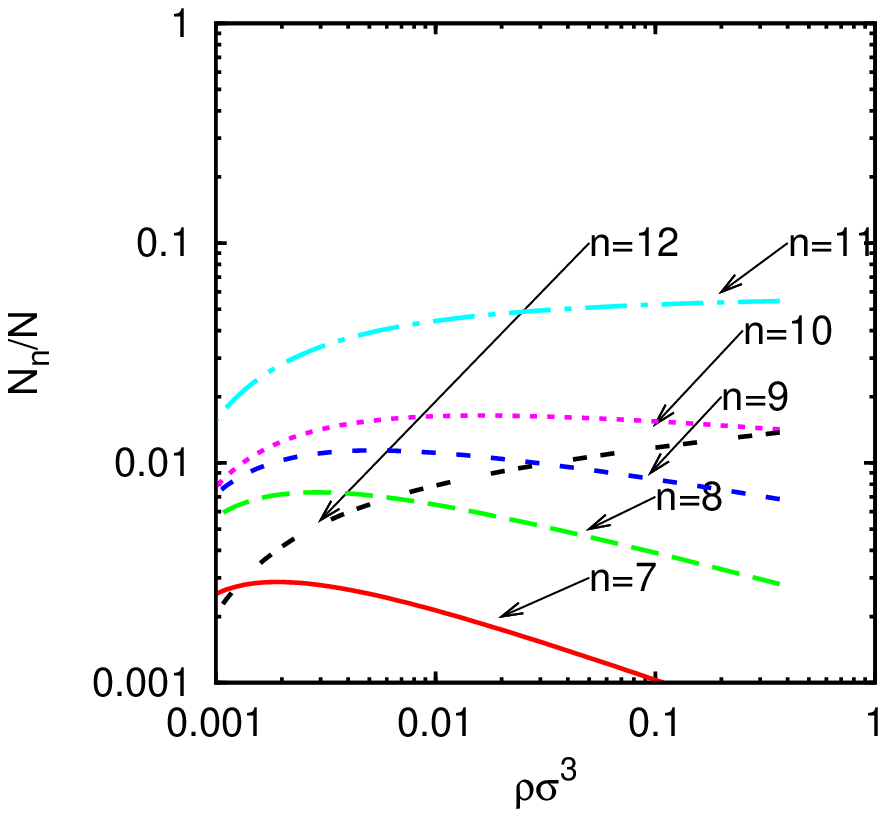} \\
\vskip1.0cm
\includegraphics[width=3.5in]{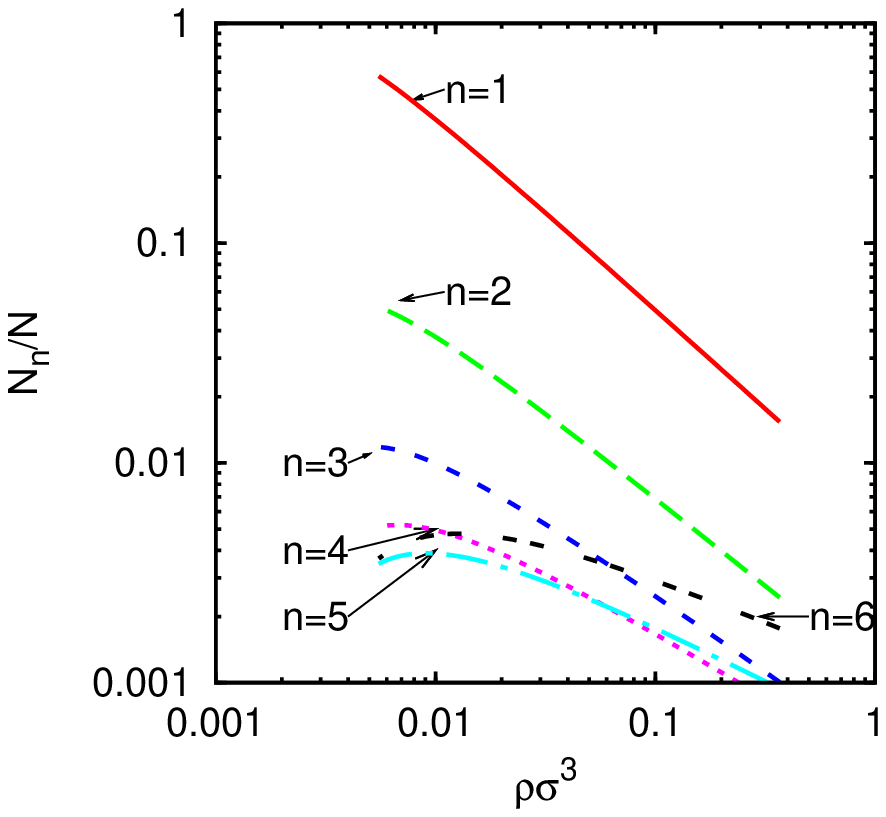} 
\hfill
\includegraphics[width=3.5in]{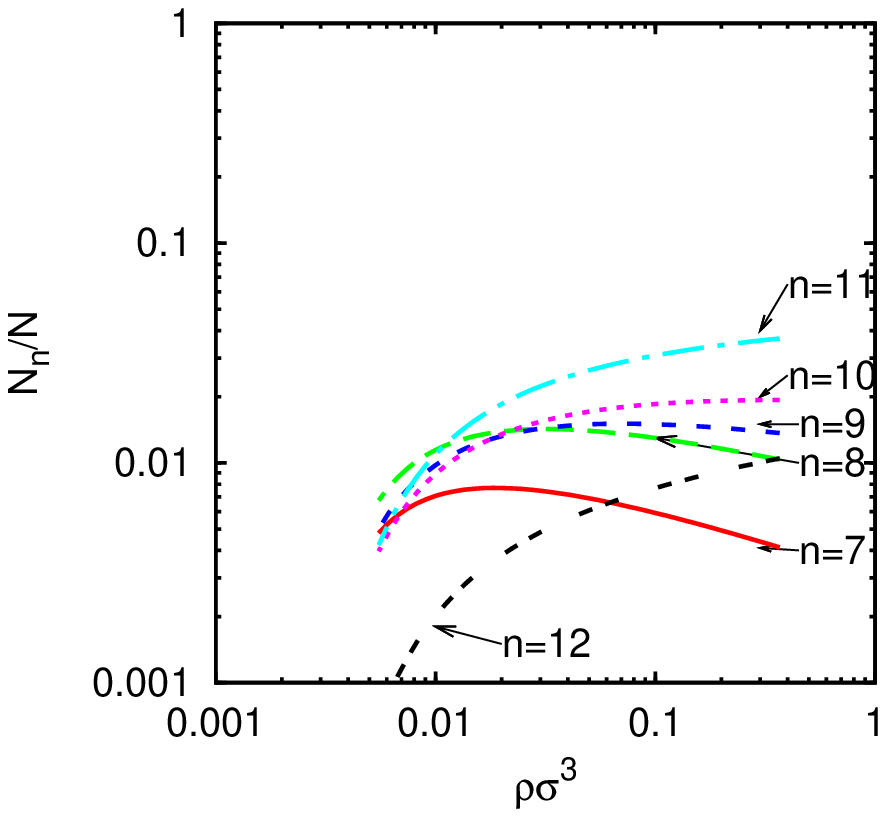} \\
\caption{Values of $N_n/N$, $n=1,2,3,\ldots,12$ as a function of the
density for $\Delta=0.5\sigma$ and $k_B T/\epsilon=0.25$ (top panels)
and $k_B T/\epsilon=0.30$ (bottom panels).In both cases curves for $n=1,\ldots, 6$
are on the left panels and those associated with $n=7,\ldots, 12$ are on the
right panels. All plots have been reported on the same scale for a better
comparison. Clusters associated with values $n=4,5$ have curves 
lying below the lower limit of 0.001 concentration  
in the case $k_B T/\epsilon=0.25$.}    
\label{fig:nval}
\end{figure}

In Fig.\ref{fig:nval} we depict the relative cluster population $N_n/N$ as a function of the reduced density
$\rho \sigma^3$ in the ideal-gas case for $n\le 12$ and two different temperatures $k_B T/\epsilon=0.25$ (top panel)
and  $k_B T/\epsilon=0.30$ (bottom panel). Temperature values were selected to bracket the expected critical
temperature $k_B T/\epsilon \approx 0.28$ on transition from a vapor phase mostly formed by monomers (at higher temperatures)
and a vapor phase with predominant clusters (at lower temperatures) in the chosen range of densities.\cite{Sciortino2010}

As expected, we observe a predominance of monomers and higher order clusters at low and high density respectively. No significant
difference is apparent for the results of the two temperatures. This is most likely due to the ideal-gas nature of the interacting
part and can be improved by using the Carnahan-Startling fluid description, as we shall see.

Next we consider the internal energy per particle $u_n=\langle U\rangle/n$ within the $n-$th cluster along with the gyration radii
defined by
\begin{eqnarray}
\label{intra:eq4}
R_g^2&=&\sum_{j=1}^{n}|\rr_j-\rr_{av}|^2/n
\end{eqnarray}
with $\rr_{av}=\sum_{j=1}^n\rr_j/n$, $\rr_j$ being the position of the
$j-$th particle. Results for both internal energy and gyration
radii for such configurations are tabulated in Table \ref{tab:1}.
This provides an additional insight on the morphologies of the obtained clusters, in particular on
the relative weak $n$ dependence of the linear size of the obtained clusters.


The results for $u_n$ as a function of temperature are reported in Table \ref{tab:2} and can be conveniently fitted by a Gaussian profile 

\begin{eqnarray} 
\label{ugauss}
u_n(T)&=& a_n \exp\left[-b_n T^2\right]+c_n~,
\end{eqnarray}
where the fitting parameters $a_n,b_n,$ and
$c_n$ for the $n=2,3,\ldots,12$ clusters ($u_1=0$ by definition) can also be found in Table \ref{tab:2}. 

From this expression we can determine the excess free energy of the cluster
$f_n^{\text{ex,intra}}=\beta F^{\text{ex, intra}}/n$ by thermodynamic integration
\begin{eqnarray}
\label{intra:eq1}
f_n^{\text{ex,intra}}\left(\beta\right)&=&\int_0^\beta dx \,u_n(1/x)
\end{eqnarray}
So that $f_n^{\text{intra}}=f_n^{\text{ex,intra,}}+f_n^{\text{id,intra}}$ with the ideal
free energy contribution being
\begin{eqnarray}
\label{intra:eq2}
f_n^{\text{id,intra}}\left(\beta\right)&=&3\ln\Lambda_n+\left(\ln
n!\right)/n-\ln v_0~, 
\end{eqnarray}
where $v_0=\pi \sigma_0^3/6$ is the volume of one $n-$cluster and 
$\Lambda_n=\sqrt{2\pi\beta\hbar^2/m_n}$ is the de Broglie thermal
wavelength, and with the excess part given by 
\begin{eqnarray}
\label{intra:eq3}
f_n^{\text{ex,intra}}&=& c_n\beta+a_n\sqrt{b_n}\left\{
\frac{e^{-b_n/\beta^2}}{\sqrt{b_n/\beta^2}}+
\sqrt{\pi}\left[\mbox{erf}\left(\sqrt{b_n/\beta^2}\right)-1\right]\right\}~.
\end{eqnarray}

The intra-cluster partition function is then
$Z_n^{\text{intra}}=v_0^ne^{-nf_n^{\text{ex,intra}}}$ (of course
$Z_1^{\text{intra}}=v_0$), where $v_0=(\pi/6)\sigma_0^3$ is the volume
of one cluster. As anticipated we here choose $\sigma_n=\sigma_0$, for all $n$, where
$\sigma_0$ is the only undetermined parameter in the theory.


\section{Additional results}
\label{sec:5}
 \subsection{Carnahan-Starling results}
In this case the theory depends upon the average diameter of a cluster $\sigma_0$. This is obtained
by the requirement that the Carnahan-Starling results best match MC results for the bulk simulations.

\begin{figure}[h!]
\begin{center}
\includegraphics[width=12cm]{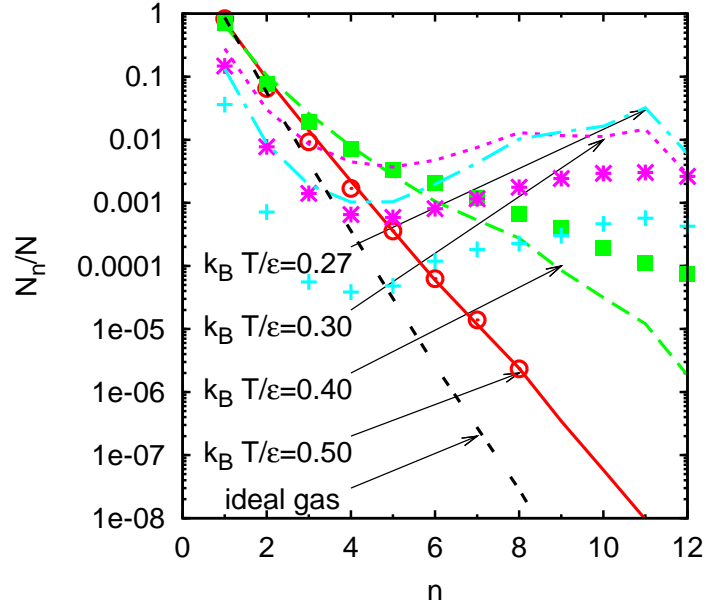}
\end{center}
\caption{Comparison between the MC data (points) and our calculations using  
the effective one component hard sphere inter-cluster partition function 
within Carnahan-Starling
approximation for $\sigma_0=2.64 \sigma$ (lines), for $N_n/N$,
$n=1,2,3,\ldots,12$ as a function of the clusters size $n$ at $\rho
\sigma^3=0.01$, $\Delta=0.5\sigma$, and various 
temperatures. Also shown is the ideal gas approximation 
}
\label{fig:Nsfit2}
\end{figure}

To this aim, we consider Monte Carlo results at $\rho \sigma^3 =0.01$ on the vapor phase, for the distribution of the cluster
sizes, with our theory. This is depicted in Fig. \ref{fig:Nsfit2} where we compare the
Carnahan-Starling approximation with the MC data for
the distribution of cluster sizes at decreasing values of temperatures starting from $k_B T/\epsilon=0.5$
which provides a good match with MC results for $\sigma_0 \approx 2.64 \sigma$.
This value is then used in all subsequent calculations.

It is important to remark that, in order to find the correct solution
for this system of equations, it 
is important to choose the one that is continuously obtained from the
solution of the ideal gas case at $\sigma_0\to 0$.

At lower temperatures the discrepancy with the MC data
for the vapor increases. This was to be expected in view of the fact that the two-layer vesiscles ($n-$clusters with $n=40$) contribution
to the vapor phase, and not included in the present computation,  becomes increasingly important \cite{Sciortino2010}.
The agreement could be clearly improved by allowing a temperature dependence of the effective cluster diameter $\sigma_0$, but
we have chosen to keep $\sigma_0$ fixed to mantain a clear control of the approximations involved in our approach.

\begin{figure}[htbp] 
\centering
\includegraphics[width=3.5in]{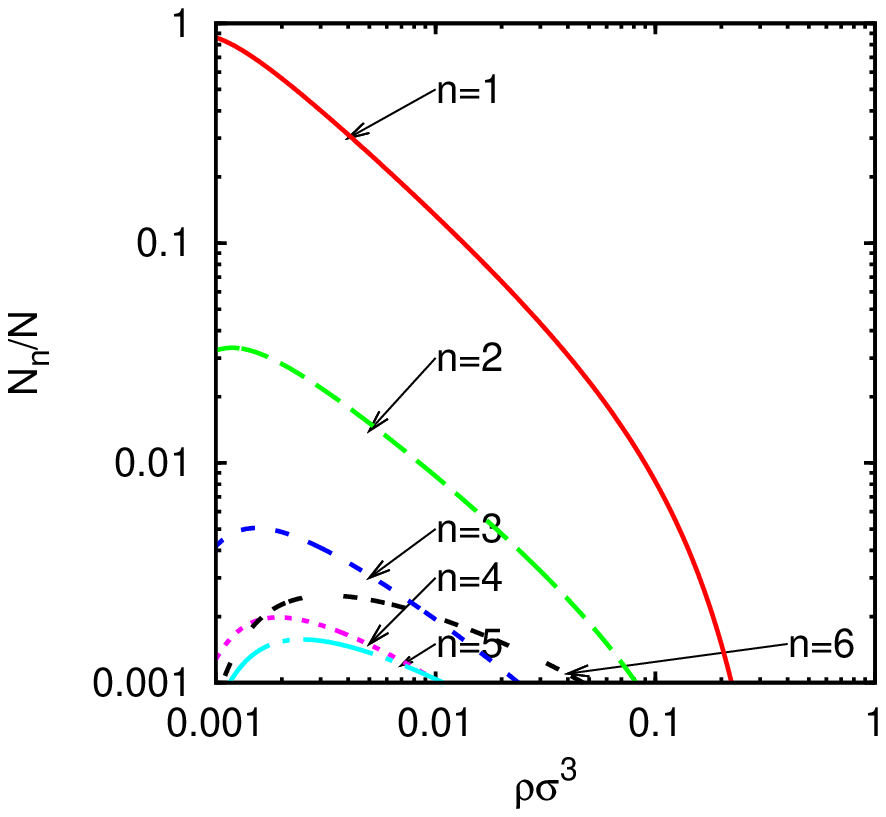}
\hfill 
\includegraphics[width=3.5in]{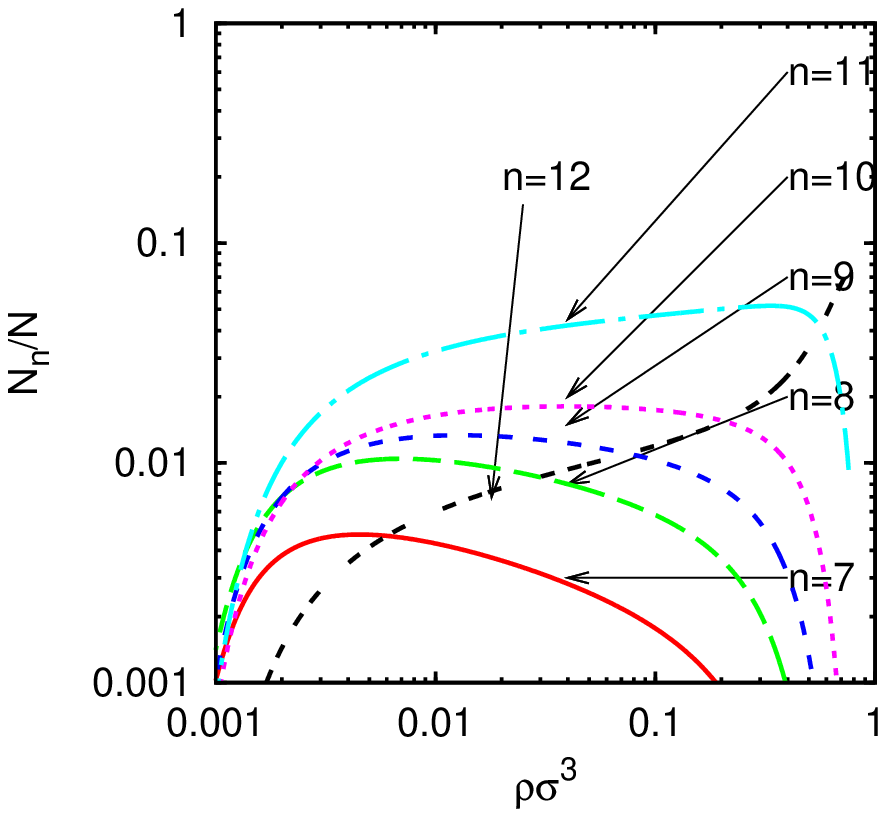} \\
\vskip1.0cm
\includegraphics[width=3.5in]{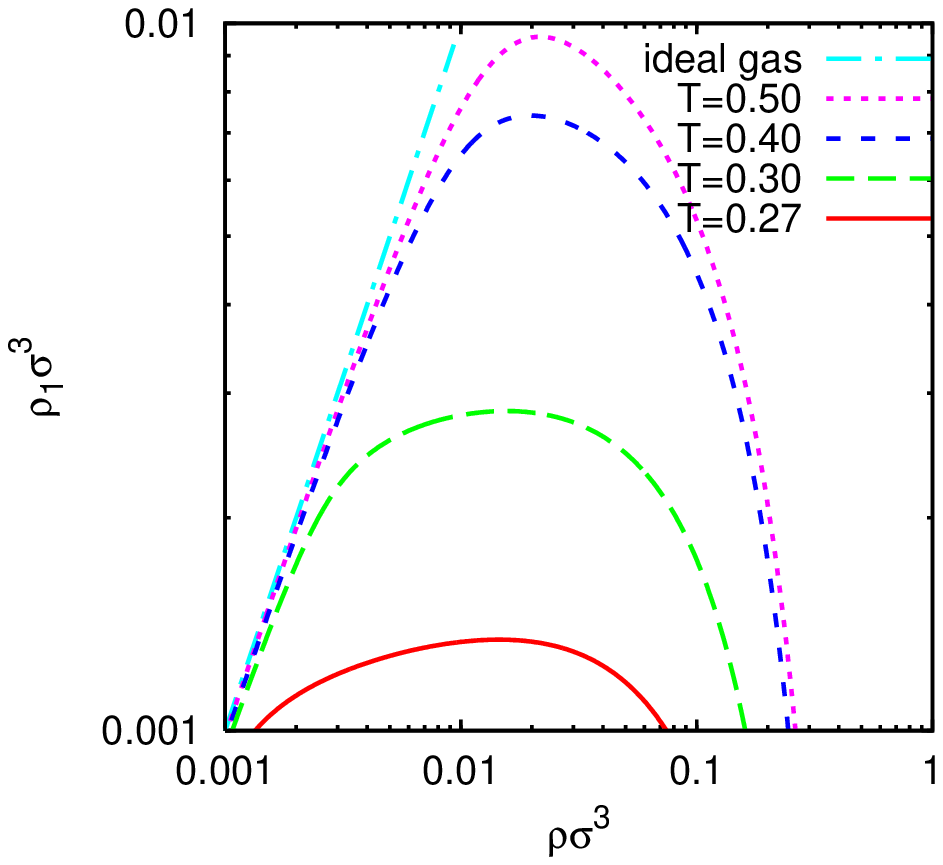} 
\hfill
\includegraphics[width=3.5in]{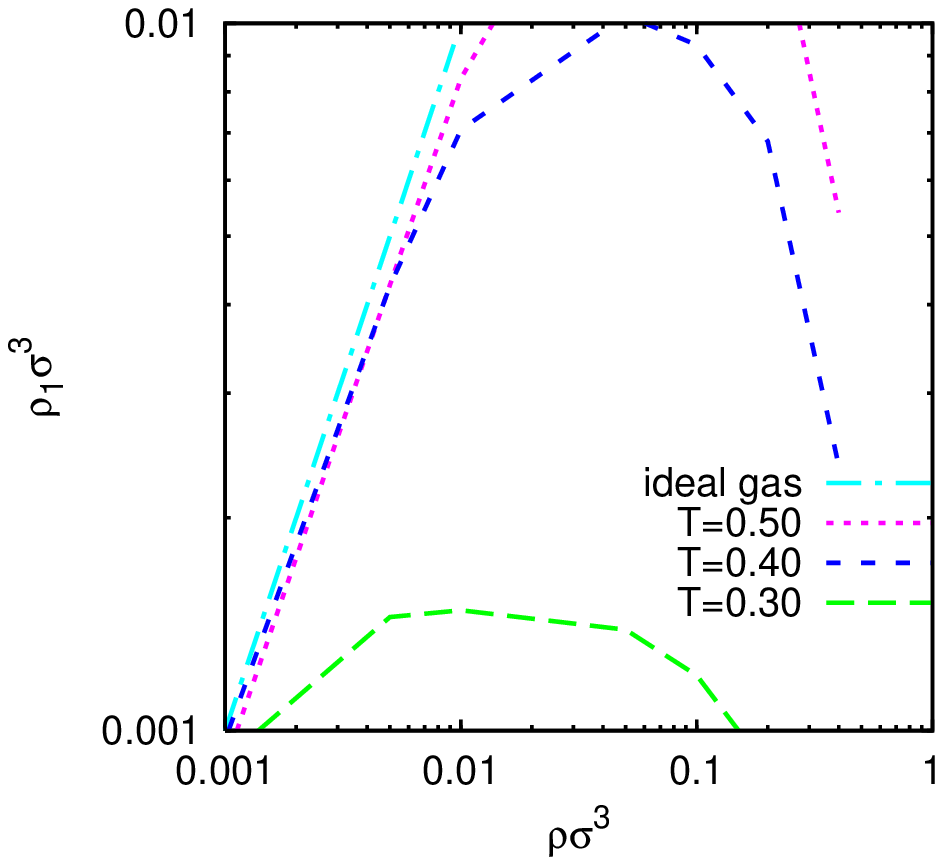} \\
\caption{Values of $N_n/N$, $n=1,2,3,\ldots,12$ as a function of the 
density for $k_B T/\epsilon=0.27$ (top panels). 
Clusters with $n=1,\ldots, 6$ are on the
left, those with $n=7,\ldots, 12$ on the right. The bottom panels
depict the monomer concentration $\rho_1 \sigma^3$ as a
function of the total density $\rho \sigma^3$ for decreasing
temperatures. The result of the present approach (left) is contrasted with MC simulations (right). 
All results refer to the $\Delta=0.5\sigma$ case with a
cluster diameter $\sigma_0=2.64 \sigma$.}     
\label{fig:nval-cs-6}
\end{figure}

In Fig. \ref{fig:nval-cs-6} (top panel ) we show the resulting cluster distribution for the $N_n/N$ as a
function of density for a temperature ($k_B T/\epsilon=0.27$ ) at the onset of the expected critical micelle concentration \cite{Sciortino2010}.
Unlike previous case with ideal gas, there is now a clear predominance of the $n \approx 10$ clusters in the whole concentration range.
Additional insights can be obtained by plotting the monomer density $\rho_1 \sigma^3$ versus the total concentration $\rho \sigma^3$
for decreasing temperatures, as reported in the bottom panels of the same Figure, where the result of the present approach
is contrasted with bulk numerical simulations of the same quantity \cite{Sciortino2010}. This clearly shows the onset of a critical concentration
where clusterization becomes the predominant mechanism at each temperature (this can be obtained by extrapolating the flat part of the
curves to the vertical exis).

\begin{figure}[h!]
\begin{center}
\includegraphics[width=12cm]{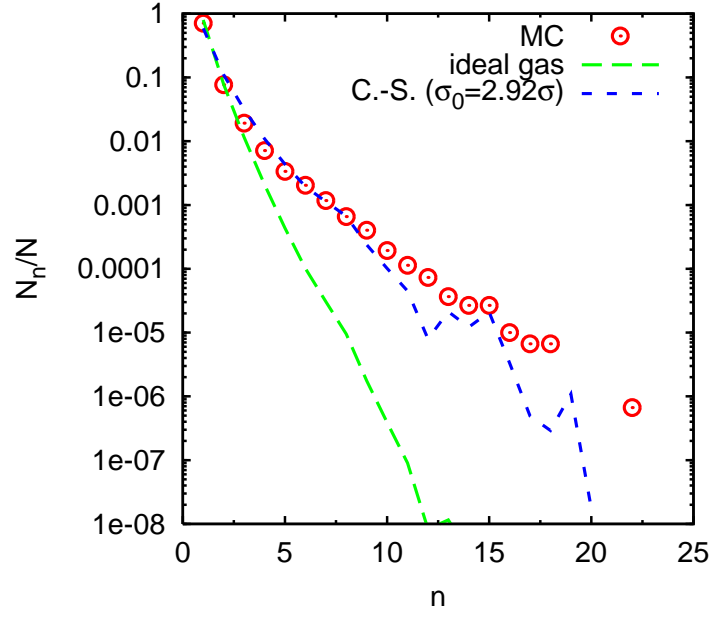}
\end{center}
\caption{Comparison between MC data and cluster theory using 
Carnahan-Starling
approximation (for $\sigma_0/\sigma=2.92$) for $N_n/N$, $n=1,2,3,\ldots,20$
as a function of the clusters size $n$ at $k_B T/\epsilon=0.4$, 
$\rho \sigma^3=0.01$, and $\Delta=0.5\sigma$.}     
\label{fig:Nsfit3}
\end{figure}

In order to assess the range of reliability of our results, we have also attempted
to include in the theory all clusters of size up to 20
particles. Fig. \ref{fig:Nsfit3} shows how the theory compares
with the MC results at $k_B T/\epsilon=0.4$ for the distribution of the cluster
sizes. Note that the vertical axis spans about $8$ order of magnitudes. Here we used a slightly different
value $\sigma_0 \approx 2.92 \sigma$ for the cluster diameters.
Our theory nicely follows the MC data for the vapor phase up to $n \leq 12$. For larger clusters discrepancies begin to
show up most likely due to the fact isolated clusters tend to frequently disaggregate during the simulation
thus providing a very low acceptance ratio. As anticipated, for this larger cluster sizes, a full simulation of the
bulk vapor phase begins to be competitive with the present methodology,  and this is the main reason
why, in the remaining of the paper, we only consider a mixture of
$n-$clusters with $n\leq 12$.  

As remarked, the present theory depends upon a free parameter (the average cluster diameter $\sigma_0$) that
is computed by a best fit with the bulk MC simulations.

\begin{figure}[h!]
\begin{center}
\includegraphics[width=12cm]{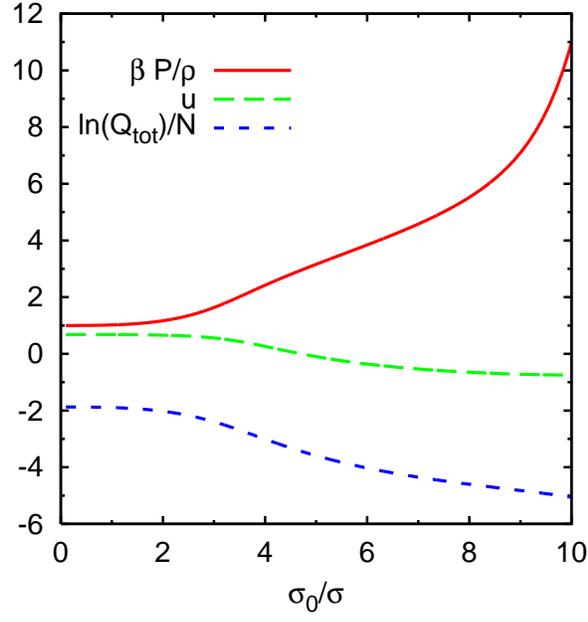}
\end{center}
\caption{Values for the compressibility factor, the internal energy
per particle, and the logarithm of the total partition function as a 
function of the $n-$cluster diameter $\sigma_0/\sigma$ at $\rho
\sigma^3=0.01$, $k_B T/\epsilon=0.5$, and $\Delta=0.5\sigma$.} 
\label{fig:thermo-cs-d}
\end{figure}

Fig. \ref{fig:thermo-cs-d} displays the sensitivity of some of the computed quantities to the choice of the average cluster diameter
$\sigma_0/\sigma$. In particular, we have considered the compressibility factor $\beta P/\rho$, the internal energy per particle
$u=U/N$ and the  reduced free energy per particle 
$\ln(Q_{\text{tot}})/N$. 
In all cases, there is a non-negligeable dependence
on the $\sigma_0/\sigma$ value indicating the importance of selecting the correct effective cluster diameter. This could be improved
by considering a distribution of cluster diameters.

\begin{figure}[h!]
\begin{center}
\includegraphics[width=12cm]{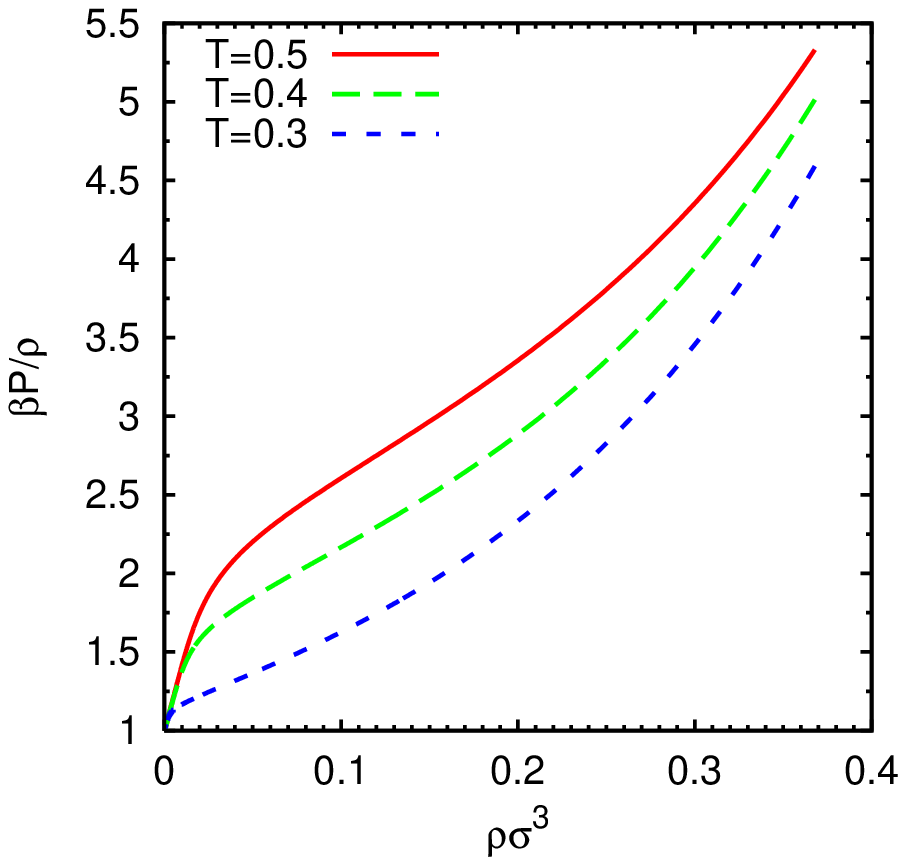}
\includegraphics[width=12cm]{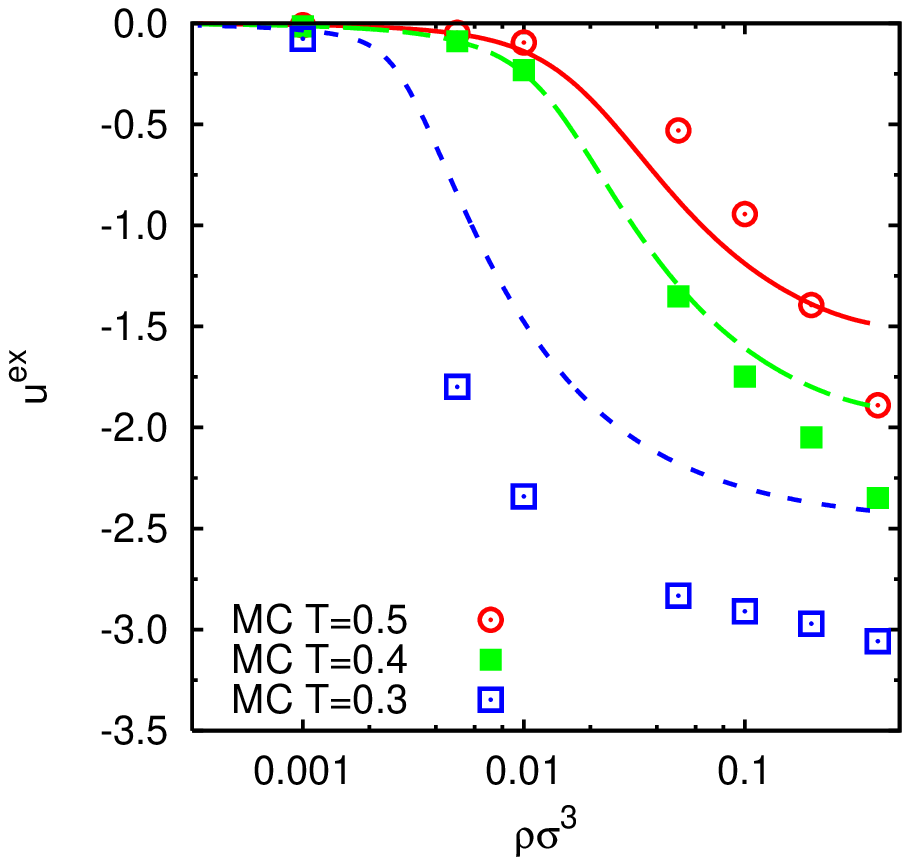}
\end{center}
\caption{Compressibility factor as predicted by the Carnahan-Starling
($\sigma_0 =2.64 \sigma$) cluster theory (top panel). On the bottom
panel we compare the MC data and the Carnahan-Starling cluster theory
(same diameter as above) for the excess internal energy per particle
for three different values of temperatures. In all cases
$\Delta=0.5\sigma$.}    
\label{fig:thermo}
\end{figure}

Notice that as
$\sigma_0$ increases the packing fraction of the clusters $\eta_{t}$ quickly exceeds unit, thus limiting the possible
range of acceptance for the cluster diameter.
Similarly, in Fig. \ref{fig:thermo} we report the compressibility factor and the
excess internal energy per particle. The excess internal energy is
compared with the MC data for the vapor phase \cite{Sciortino2009}.

\subsection{Prediction for a different range of the square well}
\label{subsec:prediction}
So far, we have considered the case where the range of interaction
(the width of the square well) $\Delta$ was $50\%$ of the particle size $\sigma$.
This is the value which has been exploited in details in past MC studies of the bulk Janus fluid \cite{Sciortino2009,Sciortino2010}
As this range decreases, typical relevant temperatures decrease and simulations become increasingly more demanding 
from the computational point of view to equilibrate. It is then no surprising that
no results have been yet reported in the literature for these ranges. On the other hand, these are the ranges most frequently encountered
in the experiments \cite{Hong2008}, and this is where the usufulness of our method can be assessed.

We have then repeated the calculations for $\Delta/\sigma=0.25$, that is half of previous value. 

\begin{figure}[htbp] 
\centering
\includegraphics[width=3.5in]{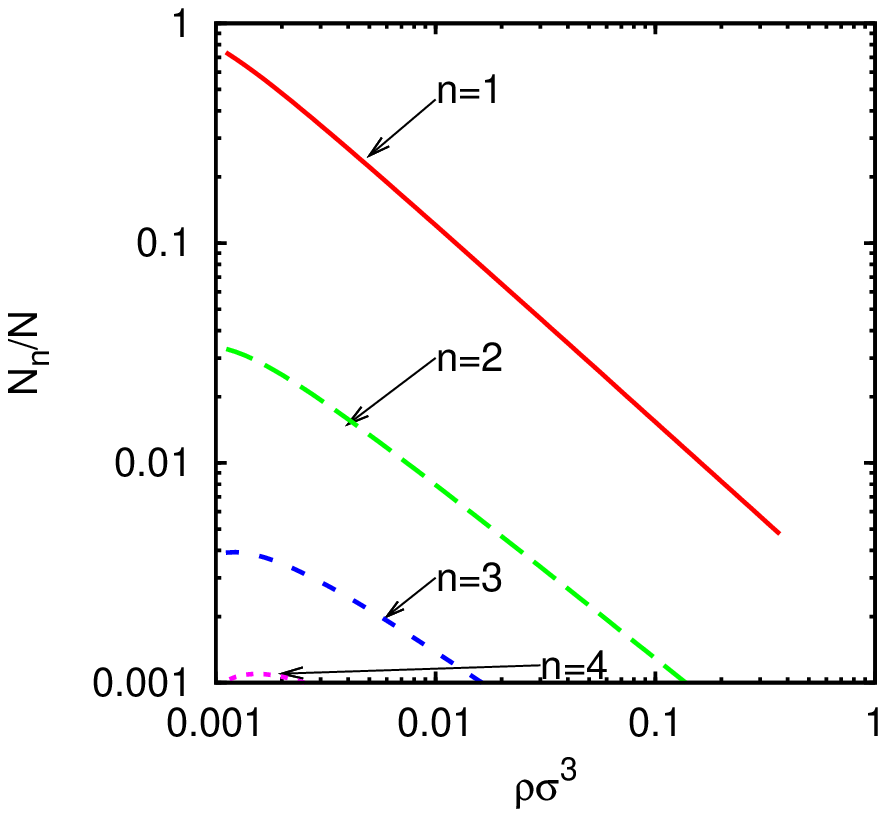}
\hfill 
\includegraphics[width=3.5in]{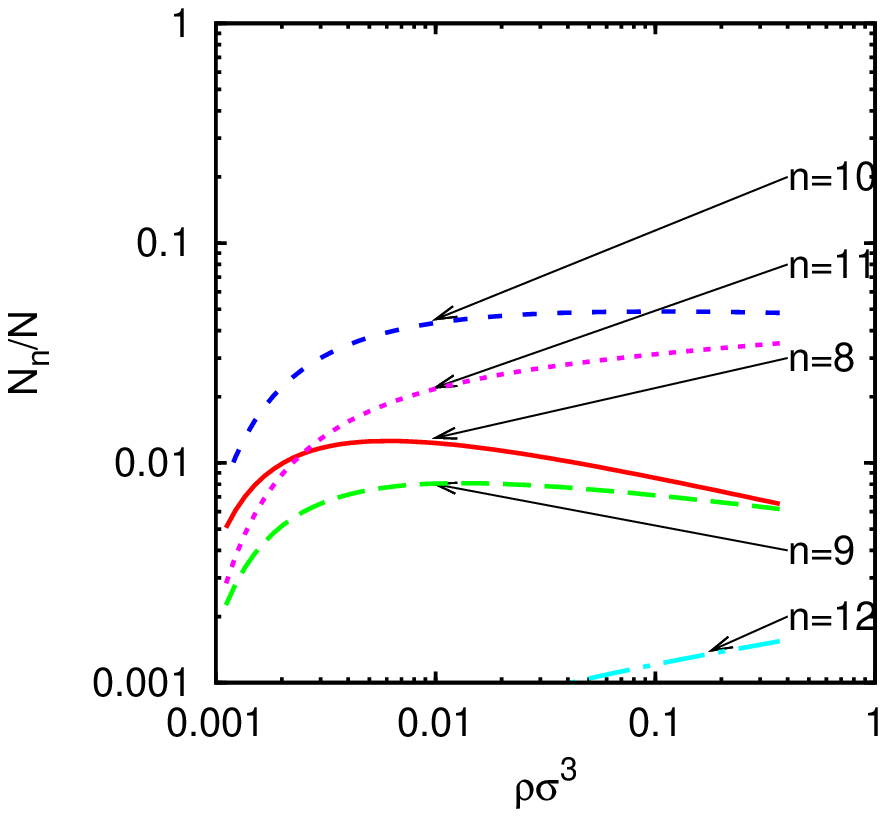} \\
\vskip1.0cm
\includegraphics[width=3.5in]{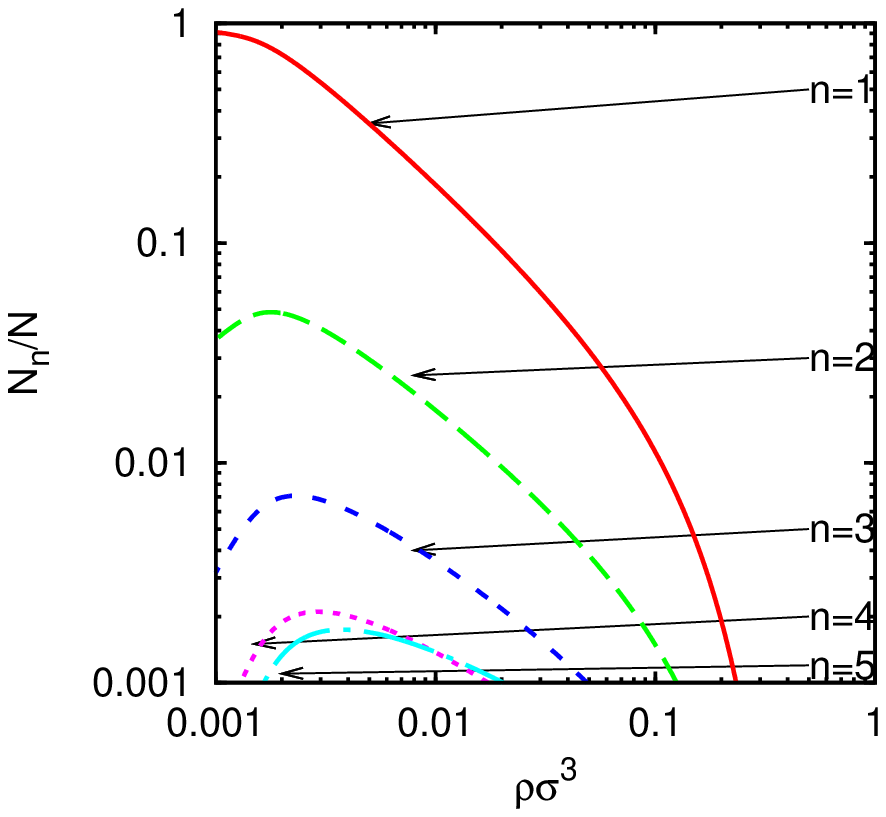} 
\hfill
\includegraphics[width=3.5in]{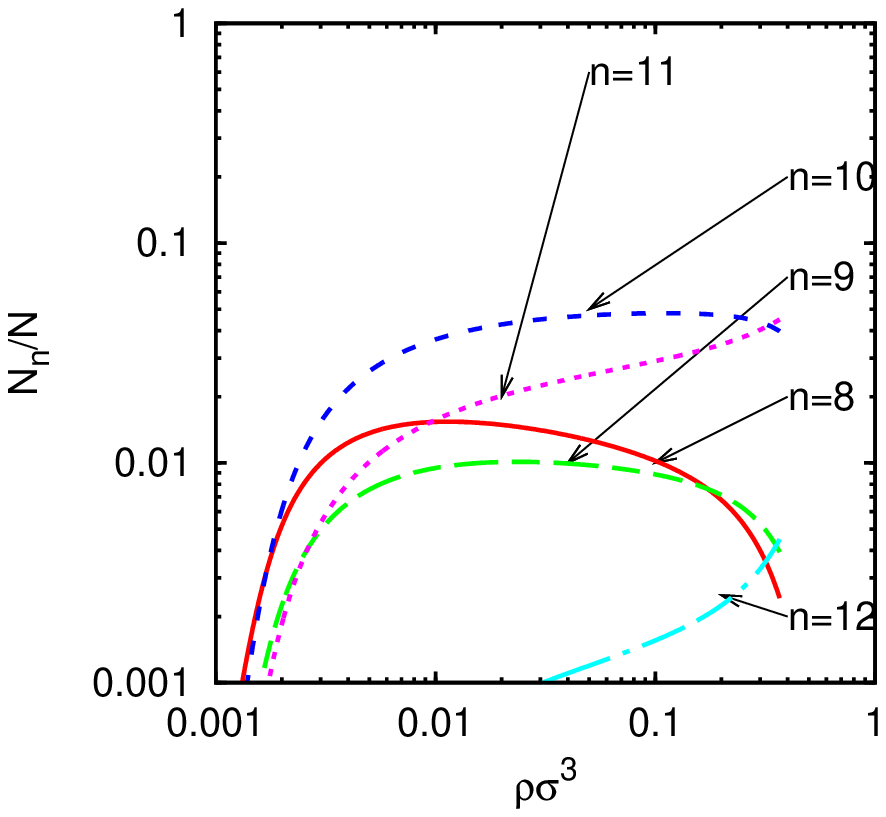} \\
\caption{Cluster distribution in the case $\Delta=0.25\sigma$. The top
panels report the ideal gas result at $k_BT/\epsilon=0.25$ ($n=1,\ldots,6$ on the left
and $n=7,\ldots,12$ on the right). This is
the same as Fig. \ref{fig:nval}. The bottom panels depict with the same distribution of curves the results obtained with  
Carnahan-Starling approximation 
at $k_BT/\epsilon=0.27$ which is the counterpart of
Fig. \ref{fig:nval-cs-6}} 
\label{fig:nval-h-0.25}
\end{figure}

Fig. \ref{fig:nval-h-0.25} 
reports the cluster distributions for the ideal and the Carnahan-Starling fluids (lower temperatures), and are the
counterpart of Figs. \ref{fig:nval} and \ref{fig:nval-cs-6}. Concentrations of the
$n-$clusters are now shifted towards higher densities with respect to the
case with the twice as wide range, as expected. Also now the role of the $10-$cluster and the $11-$cluster is inverted respect to before.
This means that lower attractive range provides, on average, smaller stable clusters, a results that can be understood on intuitive basis.
 
We also found that thermodynamic quantities considered above are only marginally affected by the reduction of the width well in the
considered range of densities and temperature.

\begin{figure}[htbp] 
\centering
\includegraphics[width=3.5in]{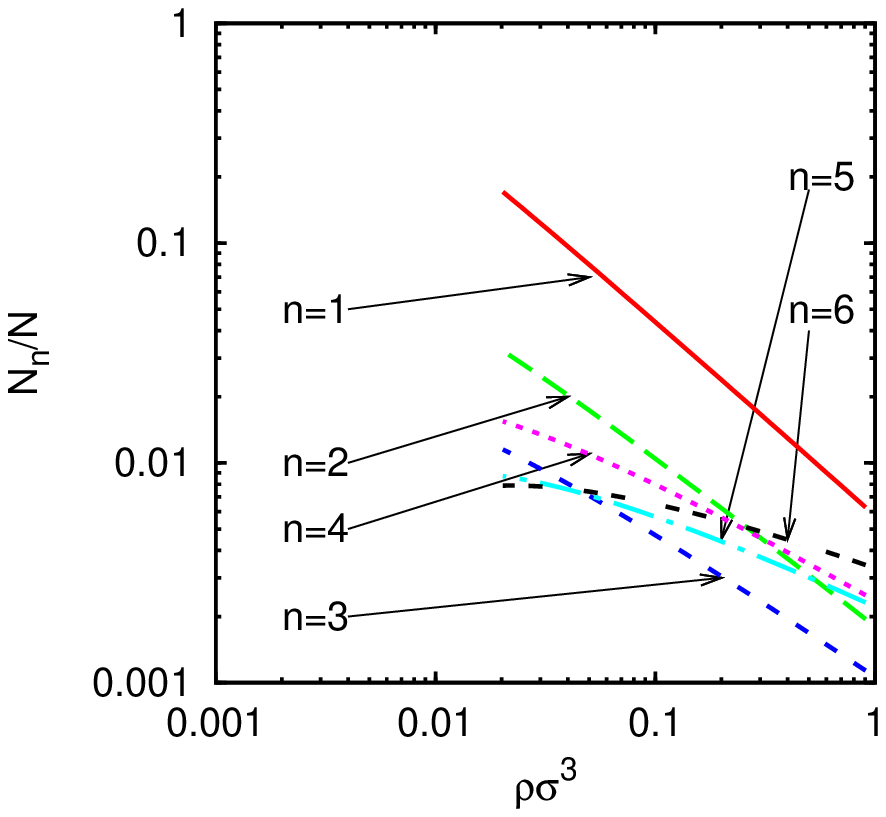}
\hfill 
\includegraphics[width=3.5in]{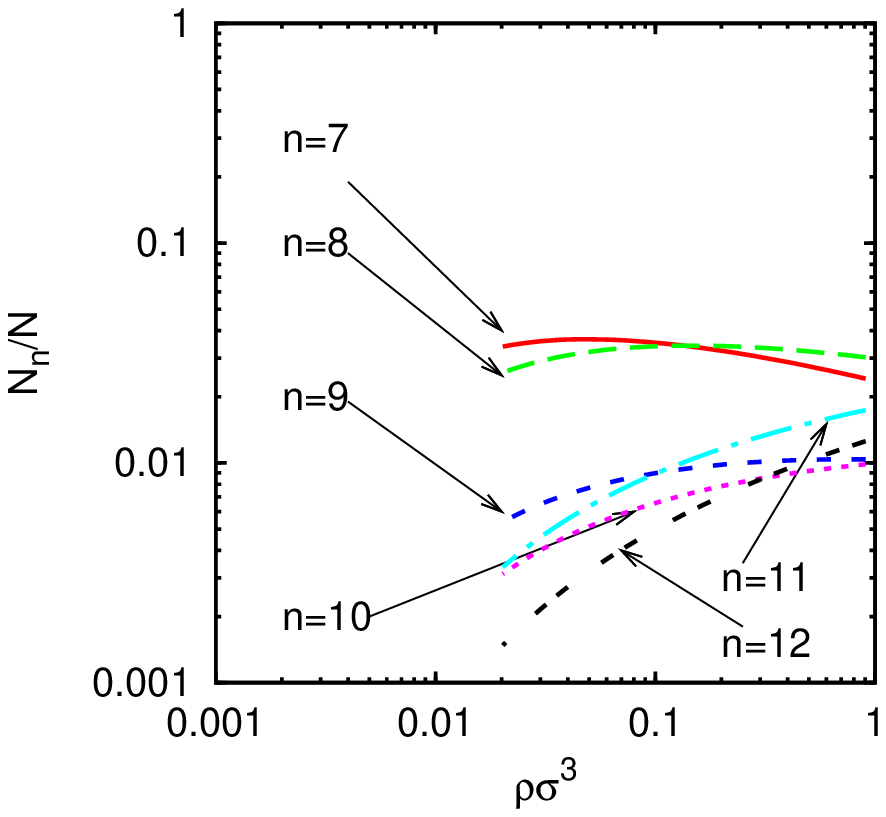} \\
\vskip1.0cm
\includegraphics[width=3.5in]{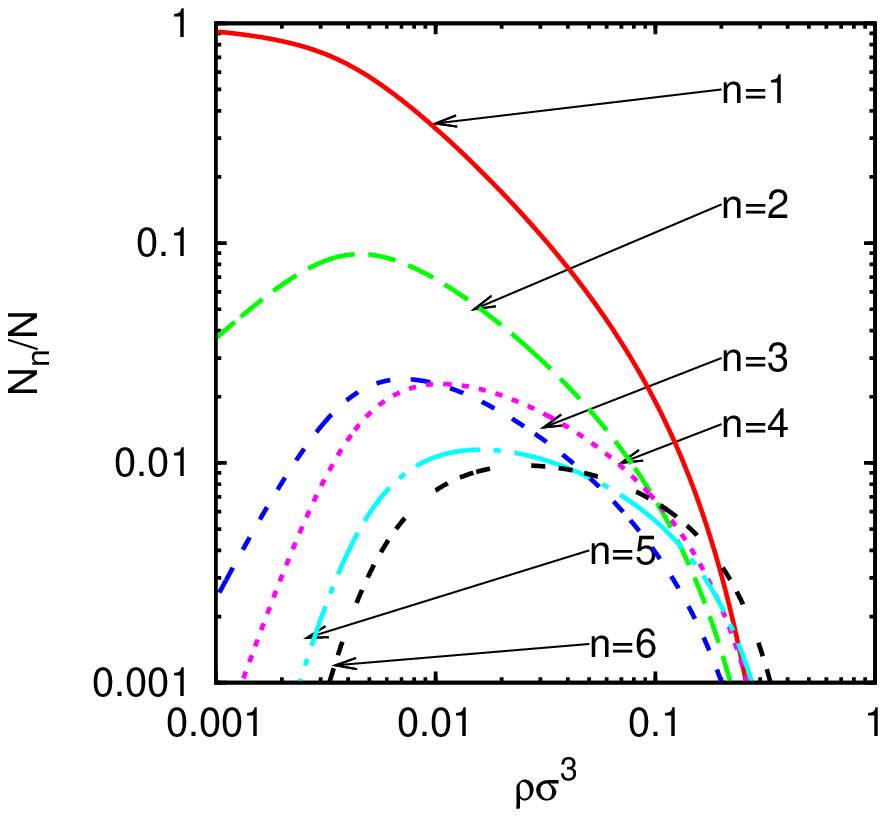} 
\hfill
\includegraphics[width=3.5in]{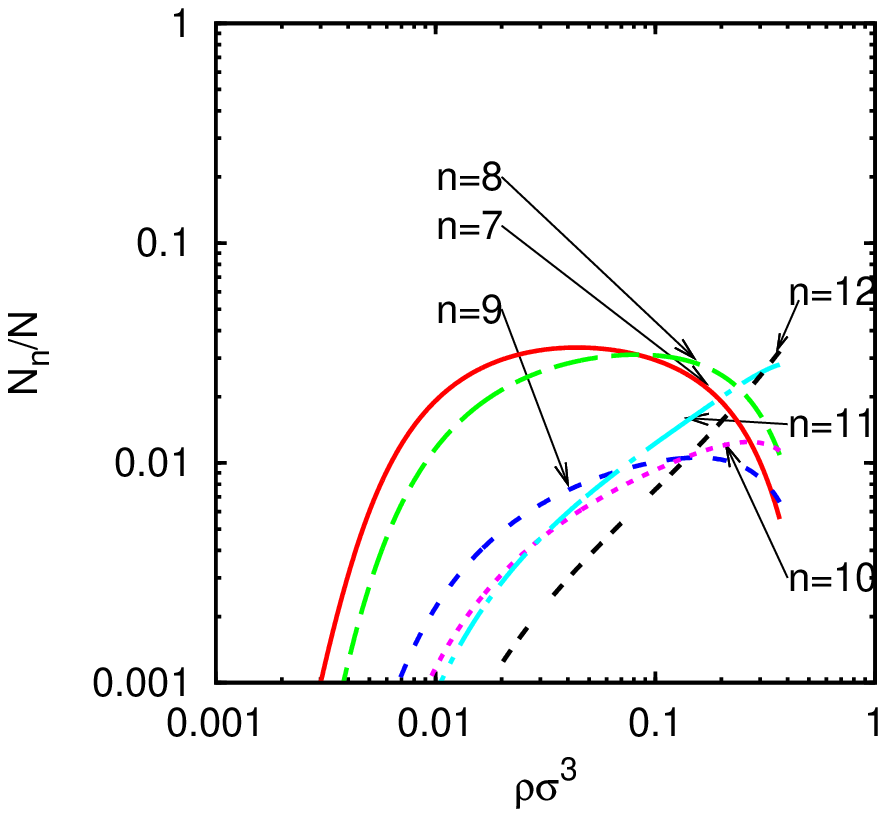} \\
\caption{Same as Fig. \ref{fig:nval-h-0.25} for $\Delta=0.15\sigma$.}
\label{fig:nval-h-0.15}
\end{figure}

We have also considered the case of $\Delta=0.15\sigma$. 
 From Fig. \ref{fig:nval-h-0.15} it is apparent that the concentrations of the
$n-$clusters are once again shifted towards higher densities respect to the
case with $\Delta=0.25\sigma$. Also now the $7-$ and $8-$clusters
seems to be the ones favoured at $T=0.27$ in a range of densities in a
neighborhood of $\rho\sigma^3=0.1$. This confirms the trend found in the case $\Delta=0.25\sigma$. 
\section{Conclusions}
\label{sec:6}
In this paper, we have constructed a cluster theory for the vapor of Janus fluid. This is
an approach that is complementary to previous studies based on highly demanding MC simulations \cite{Sciortino2009,Sciortino2010},
with the aim of providing a detailed description of the vapor phase in view of its remarkable unusual micellization
properties. 

The main idea behind the present approach is to consider the vapor phase as formed by clusters, containing an increasing number of
particles, that are weakly interacting among each other so that simple fluid models -- such as ideal gas or hard spheres -- can
be used to mimick their physical properties. The internal degrees of freedom of each clusters are instead obtained through a direct
MC simulation of a single isolated cluster, a much simpler task as compared to the bulk simulation and a procedure
akin to those used in the framework of simple fluids \cite{Tani83} is then used to combine the two calculations and obtain the
full description of the system.

It is worth noticing that, in the ideal-gas case, a similar procedure has also been already implemented in micellization
theories by several groups \cite{Nagarajan91,Israelachvili76}, and the results we obtain in the present context are
quite consistent with those.

There are two basic reasons why we expect this approach to be valuable. First because previous full bulk simulations
showed micelles to be only weakly interacting in the vapor density range and hence a simple description for the
inter-cluster part is expected to be sufficient. Second, because it has been observed that the vapor properties are mostly dominated
by particular cluster sizes corresponding to $n \approx 10$ and $n \approx 40$ particles, so only a limited number of
cluster sizes is necessary to obtain a complete description.
 
In the present work, we have considered clusters up to $12$ particle and compared the ideal-gas description with
the description of a gas of hard-sphere spheres, mimicking the original clusters and with an effective cluster diameter $\sigma_0$,
using the Carnahan-Starling approximate description. The value of $\sigma_0$ has been obtained by a matching of the
results for the internal energy with full bulk MC simulations. A good agreement was found at $k_B T/\epsilon=0.5$ and
at densities $\rho \sigma^3=0.01$ when $\sigma_0 \approx 2.64 \sigma$. Results from  Carnahan-Starling theory is found to be far superior
as compared to the ideal-gas description, thus emphasizing the importance of inter cluster correlations in the vapor phase. 
 
We also considered higher sizes clusters (of up to $20$
particles) but the agreement with the simulations for the larger sizes becomes less satisfactory. 
The theory becomes less and less accurate as
oscillations in the behavior of the concentrations of the big
clusters with size appear. This may be due to the difficulty in an accurate
determination of the internal energy of isolated big clusters. On this
respect in order to be able to observe the vesicles (clusters
of around $40$ particles \cite{Sciortino2009}) phenomenology we
certainly need to include additional insights to avoid the task of the solution of a system of about $40$ coupled equations. An additional
difficulty consists in the fact that in this case the single diameter effective approximation used for all clusters up to $12$ in the present
study, will no longer be realistic, not even at the simplest possible level of description. Both these problems could be tackled
by focussing only on clusters bracketing the interesting ones ($n \approx 10$ and $n \approx 40$ in the present case).

We showed that in accord with the simulation results of Ref.
\onlinecite{Sciortino2009}, at temperatures around $k_B T/\epsilon=0.27$ there is a gap of densities where
the number of clusters of $11$ particles (micelles) surpasses the
number of any other cluster. This gap shrinks as we increase the temperature.

The determined approximation to the partition function of the vapor
phase of the Janus fluid can then be used to compute various
thermodynamical quantities.

We found reasonable quantitative agreement between the Monte Carlo data of
Ref. \onlinecite{Sciortino2009} and our theory for the excess internal energy of
the vapor phase of the Janus fluid. We additionally computed the compressibility factor
for which no simulation data are yet available.

Having validated the model against numerical predictions for $\Delta=0.5 \sigma$ we pursued the analysis for lower
widths of the well, values that are closer to the experimental range of interactions \cite{Hong2008}. In view of the overall 
decrease in the attractions, characteristic critical temperatures also decrease, thus making
numerical simulations increasingly demanding from the computational point of view. 

For the case $\Delta=0.25 \sigma$ we produced new predictions for the concentrations, the compressibility factor, and
the internal energy per particle as a function of density. In
particular we saw that as the range of the attraction diminishes the
Janus fluid prefer to form cluster of a lower number of particles.

Consistent results are also found for the case $\Delta=0.15 \sigma$, 
a value which rather close to those used in experiments.

An attempt to push the cluster theory to bigger cluster sizes showed
that the theory becomes less and less accurate as
oscillations in the behavior of the concentrations of the big
clusters with size appear. This may be due to the difficulty in an accurate
determination of the internal energy of isolated big clusters. On this
respect in order to be able to observe the vesicles (clusters
of around $40$ particles \cite{Sciortino2009}) phenomenology we
certainly need to include additional insights to avoid the task of the solution of a system of about $40$ coupled equations. An additional
difficulty consists in the fact that in this case the single diameter effective approximation used for all clusters up to $12$ in the present
study, will no longer be realistic, not even at the simplest possible level of description. Both these problems could be tackled
by focussing only on clusters bracketing the interesting ones ($n \approx 10$ and $n \approx 40$ in the present case).

Two additional perspectives will be the subject of a future study. First the dependence on coverage $\chi$ 
could also be tackled using the present approach, and this would provide invaluable information
on the micellization mechanism for small coverage, a task that is still our of reach to direct numerical simulations.
Secondly, it would be extremely interesting to address the issue of the reentrant phase diagram
and the (possible) existence of an additional liquid-liquid critical point. 
This has been recently attempted in a very recent preprint \cite{Reinhardt10},using a monomer-cluster equilibrium
theory in the same spirit of that presented here.


\begin{acknowledgments}
RF would like to acknowledge the support of the National
Institute of Theoretical Physics of South Africa. AG acknowledges the support of a PRIN-COFIN 2007B58EAB grant.
FS acknowledges support from  ERC-226207-PATCHYCOLLOIDS and  ITN-234810-COMPLOIDS.
\end{acknowledgments}

\bibliographystyle{apsrev}


\clearpage

%
\clearpage
\clearpage
%
\clearpage
\clearpage
\clearpage
\clearpage
%
\clearpage

\clearpage
\begin{table}[h]
\begin{tabular}{cccc}
\hline
$n$ & $\langle U\rangle/n$ & $U$ & $R_g$\\
\hline
1       &0      &0      & 0\\
2       &-0.5   &-1     & $\sim 1/2$\\
3       &-1     &-3     & $\sim 1/\sqrt{3}$\\
4       &-1.5   &-6     & 0.83\\
5       &-2.0   &-10    & 0.76\\
6       &-2.50  &-15    & 0.75\\
7       &-2.71  &-19    & 0.91\\
8      	&-2.88	&-23    & 0.93\\
9	&-3.10	&-28    & 0.96\\
10	&-3.20	&-32    & 1.00\\
11	&-3.36	&-37    & 1.04\\
12	&-3.42	&-41    & 1.08\\
\hline 
\end{tabular}
\caption{The low temperatures internal energy per particle of the
clusters with up to 12 particles when $\Delta=0.5\sigma$. Also shown
is the gyration radius $R_g$ defined in Eq.(\ref{intra:eq4})
.} 
\label{tab:1}
\end{table}
\clearpage
\begin{table}[h]
\begin{tabular}{ccc||cc||cc||c}
\hline
&& $\Delta=0.5\sigma$ && $\Delta=0.25\sigma$ && $\Delta=0.15\sigma$&\\
\hline
$n$ & $a_n$ & $b_n$ &  $a_n$ & $b_n$ & $a_n$ & $b_n$ & $c_n$\\
\hline
2&         0&         1&       	0&      1&      0&        1&     -0.50\\
3&        -0.337&   3.880&  -0.339&  6.905&  -0.346&   10.780&   -0.67\\
4&        -0.778&   4.670&  -0.771&  7.502&  -0.774&    7.975&   -0.75\\
5&        -1.226&   5.162&  -1.025&  5.890&  -1.034&    9.366&   -0.80\\
6&        -1.700&   5.600&  -1.381&  7.361&  -1.207&    9.214&   -0.83\\
7&        -1.899&   5.263&  -1.423&  6.767&  -1.480&    8.277&   -0.86\\
8&	  -2.064&   5.080&  -1.520&  4.179&  -1.551&    8.503&   -0.88\\
9&        -2.301&   5.478&  -1.579&  4.367&  -1.681&   10.160&   -0.89\\
10&	  -2.394&   5.509&  -1.725&  4.271&  -1.551&    9.419&   -0.90\\
11&       -2.556&   5.644&  -1.846&  4.829&  -1.696&    9.755&   -0.91\\
12&	  -2.598&   6.077&  -1.854&  5.723&  -1.814&   10.567&   -0.92\\
\hline 
\end{tabular}
\caption{Fit to a Gaussian of the energy per particle as a function of
the temperature (see Eq. (\ref{ugauss}))}. $c_n$ values are common to the three cases.
\label{tab:2}
\end{table}
\end{document}